\documentclass[10pt,conference]{IEEEtran}
\usepackage{amsmath,amssymb,amsfonts}
\usepackage{color,soul}
\usepackage{mathtools}
\usepackage{amsmath}

\usepackage[export]{adjustbox}
\usepackage{caption}
\usepackage{url}
\usepackage{ifpdf}
\usepackage{subfig}
\usepackage{lipsum}
\usepackage{multicol}
\usepackage{wrapfig}
\usepackage{todonotes}
\usepackage{algorithmic}
\usepackage{lettrine}
\usepackage{amsmath}
\usepackage{graphicx}
\usepackage{textcomp}
\usepackage[english]{babel}
\usepackage{amsthm}
\usepackage[ruled,vlined]{algorithm2e}

\theoremstyle{definition}

\DeclareGraphicsExtensions{.png,.pdf,.jpg,.svg}
\usepackage{array}
\usepackage{multirow}
\usepackage{multicol,tabularx,capt-of}
\usepackage{multirow}
\usepackage{cite}
\usepackage{xcolor}
\usepackage{array,multirow,textcomp}
\theoremstyle{definition}

\theoremstyle{remark}
\usepackage{authblk}

\usepackage{float}
\usepackage{geometry} 
\usepackage{authblk}
\usepackage{etoolbox}

\begin{document}
\title{Intent Profiling and Translation Through Emergent Communication}
\author[$*$]{Salwa Mostafa}
\author[$+$]{Mohammed S. Elbamby}
\author[$+$]{Mohamed K. Abdel-Aziz}
\author[$*$]{Mehdi~Bennis}
\affil[$*$]{Centre for Wireless Communications, University of Oulu, FI-90014 Oulu, Finland.}
\affil[$+$]{Nokia Bell Labs,  Espoo, Finland}
\affil[$*$]{salwa.mostafa, mehdi.bennis@oulu.fi}
\affil[$+$]{mohamed.3.abdelaziz, mohammed.elbamby@nokia-bell-labs.com}
\maketitle

\begin{abstract}

To effectively express and satisfy network application requirements, intent-based network management has emerged as a promising solution. In intent-based methods, users and applications express their intent in a high-level abstract language to the network. Although this abstraction simplifies network operation, it induces many challenges to efficiently express applications' intents and map them to different network capabilities. Therefore, in this work, we propose an AI-based framework for intent profiling and translation. We consider a scenario where applications interacting with the network express their needs for network services in their domain language. The machine-to-machine communication (i.e., between applications and the network) is complex since it requires networks to learn how to understand the domain languages of each application, which is neither practical nor scalable. Instead, a framework based on emergent communication is proposed for intent profiling, in which applications express their abstract quality-of-experience (QoE) intents to the network through emergent communication messages. Subsequently, the network learns how to interpret these communication messages and map them to network capabilities (i.e., slices) to guarantee the requested Quality-of-Service (QoS). Simulation results show that the proposed method outperforms self-learning slicing and other baselines, and achieves a performance close to the perfect knowledge baseline.

\end{abstract}

\begin{IEEEkeywords}
Intent-based networking, network automation, emergent communication, network slicing, multi-agent reinforcement learning.
\end{IEEEkeywords}

\section{Introduction}

The proliferation of various services and applications in 5G and beyond networks, such as Augmented/Virtual Reality (AR/VR),  cloud gaming, Vehicle-to-everything (V2X) communication, and smart industry, drives network service providers to move toward automated network and service management. The reason is that traditional manual configuration and management cannot support the stringent and diverse demands of services and applications. Intent-based networking (IBN) introduces a simple and efficient autonomic and autonomous way to configure and manage networks~\cite{leivadeas2022survey}. IBN relies on understanding what network users and applications want and what network operators can offer to optimize the alignment of the network operations with service or application needs. The application/user needs are expressed in an abstract and high-level language, whereby IBN focuses on understanding them instead of configuring the network operations. Therefore, {\em the intent is defined as a high-level and abstract description of the network services}~\cite{clemm2020intent}.

On the other hand, to facilitate the realization of service requirements for diverse 5G use cases and allow flexible network operation and management, network capabilities are offered with guaranteed network quality-of-service (QoS) levels, for example using network slicing ~\cite{afolabi2018network}.   Network slicing allows the creation of multiple end-to-end logical networks on a shared physical and virtual infrastructure to allow the separation of different network traffic. Each logical network (i.e., network slice) supports a certain QoS or network functions through a network slice template. The network slice template installs a network workflow and functions based on the service user's requirements or intents.  

\begin{figure}
    \centering
    \includegraphics[width = 3 in, height = 4 in]{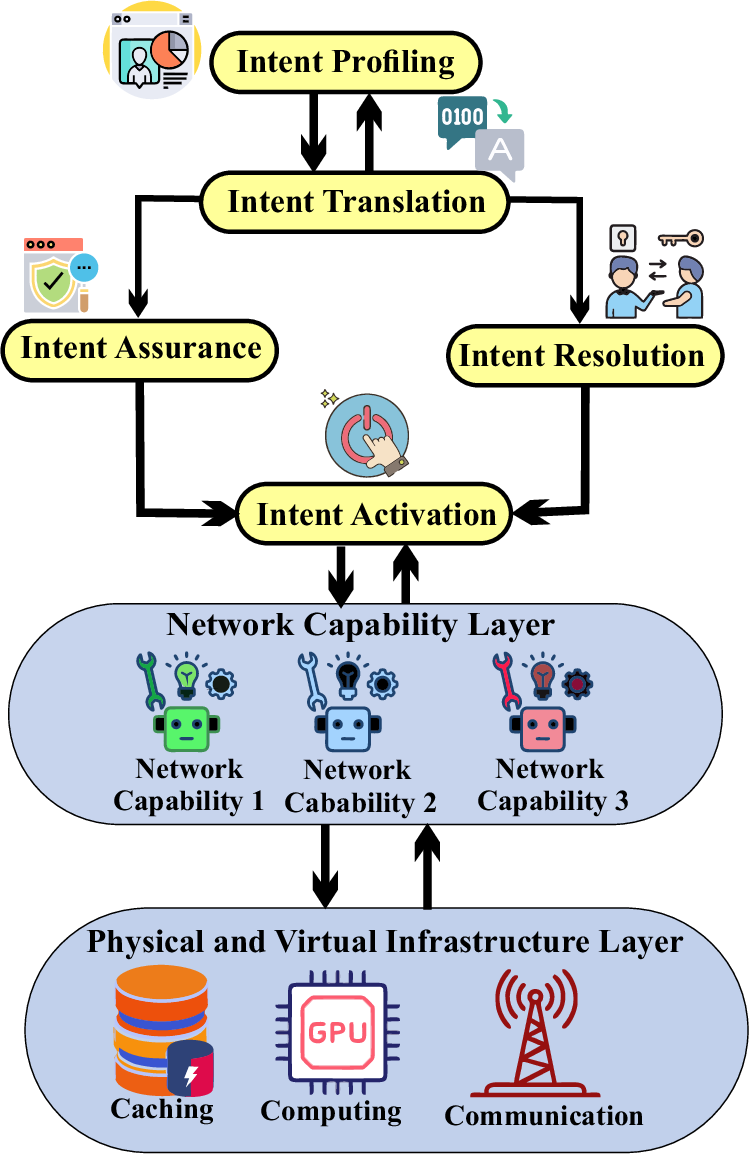}
    \caption{Interaction of the main IBN components.}
   \label{Fig1}
\end{figure}

IBN provides a complete life cycle to the intent or requested service, which takes place over five main steps to form a closed-loop automation
(CLA). The five steps are intent profiling, intent translation, intent resolution, intent activation, and intent assurance~\cite{clemm2020intent}, as shown in Fig~\ref{Fig1}. The first step is intent profiling, where the user interacts with the network and they collaborate towards expressing a meaningful intent for the network. (i.e., what the user expects as an outcome from the network or service). The second step is intent translation, where the expressed intent is converted into network policy and low-level configuration to the network functions and devices. The third step is intent resolution, which solves the potential conflict between independently submitted intents. The fourth step is intent activation, which activates the network functions and services to provide the intended customized service. The fifth step is intent assurance, which indicates the success of the deployed intent in the network throughout its dynamic life cycle. In this work, we focus on two main fundamental steps in IBN within a machine-to-machine interaction scenario, which are intent profiling and intent translation. 

\subsection{Related Work}

Internet Engineering Task Force (IETF) classified the intent profiling based on the network type, the intent scope, and the network scope. Intent profiling has been investigated in~\cite{leivadeas2021vnf,kim2020ibcs,rafiq2020intent,aklamanu2018intent,aklamanu2019intent}. The studies in~\cite{leivadeas2021vnf,kim2020ibcs,rafiq2020intent} proposed human-friendly interfaces such as graphical user interface (GUI) and templates, where network users choose, from a drop-down menu and template filling choices, what they want in their requested services. The deployment of network slices in 5G networks has led to the use of a generic network slice template, where network users express their intents through a set of attributes that refer to network performance metrics~\cite{aklamanu2018intent,aklamanu2019intent}. Although GUI and templates are human-friendly tools to express intents, they do not allow network users to express demands not supported as a possible option. Thus, the studies in~\cite{kiran2018enabling,scheid2020controlled} proposed natural language processing (NLP)-based intents, where network users write down or orally express what they expect from a network service in a human language, then the network interprets and translates it to a network configuration. However, NLP requires a specific grammar format so that important information can be detected and a correct intent is expressed. The work in~\cite{riftadi2019p4i,jacobs2018refining} proposed an intent-based language such as Nile language to express the intent in readable and abstracted technical details. The work in~\cite{xia-sdnrg-nemo-language-04} proposed a domain-specific language named NEtworking MOdeling
(NEMO), which declares intent with information about network services and resources. However, the intent-based language requires technical users (i.e., network operators/
administrators), which makes it not general enough for all types of applications.

After intent profiling, the received high-level intent must be translated into low-level network policy that can be easily rendered into network configuration scripts. The studies in~\cite{riftadi2019p4i,leivadeas2021vnf,tuncer2018northbound,alsudais2017hey,toy2021intent,borsatti2019intent} proposed several translation methods based on the intent expression type and scope. The work in~\cite{riftadi2019p4i,leivadeas2021vnf} proposed a template/blueprint-based translation, which relies on pre-defined configuration files that contain the main configuration set up to network devices and functions. It also has some variables that can be modified according to the intent, which makes it easier to use with GUI/template intent profiles. Unfortunately, a blueprint/template-based translation scheme is efficient only when there is a single mapping between the intent to a network policy or multiple network policies, but due to the abstract level of the intent, this is not necessarily the case. Thus, a mapping mechanism is proposed in~\cite{tuncer2018northbound} as another way to map multiple received intents at the same time to prevent subsequent conflicts in terms of the number of functions needed. Moreover, the network service descriptors (NSD)--based translation, proposed in~\cite{borsatti2019intent}, contains deployment templates that are directly used by an orchestrator to manage, configure, and deploy a network service. Another translation scheme is the keyword-based translation proposed in~\cite{alsudais2017hey,toy2021intent}. It identifies the associated keywords with each intent and maps them to specific rules or template policies. Nevertheless, the traditional mapping and translation mechanisms mentioned above do not guarantee an accurate translation. 

Despite the significant effort in academia and industry for expressing and translating intents, most of the aforementioned work focused on human-to-machine scenarios and no attention has been paid to machine-to-machine scenarios. Different from the human-to-machine interaction, where humans can articulate their needs through verbal, GUI, and drop-down menus, machine-to-machine interactions require learning a common language (i.e., non-verbal) for applications to express their intents. Moreover, existing literature assumes intents are easily expressed as network QoS requirements by applications. However, providing services that satisfy stringent application requirements means applications have to be able to express their needs in their domain language. This condition is complex since it requires networks to learn how to interpret these domain languages, which is not practical nor scalable. Moreover, confining intents to pre-defined lists of generic intents that the network can understand restricts the potential of future networks’ ability to support the variety of applications, each having its own needs/goals. 

Therefore, in this work, we propose a simple and flexible intent profiling framework for machine-to-machine interaction that leverages a set of communication messages that a machine learns how to associate with different application intents. Moreover, to tackle the drawbacks of traditional mapping and translation mechanisms, we propose a mapping technique that relies on artificial intelligence (AI) that is able to learn and improve translation over time through experience. The network translates the received communication messages to configured network slices that can support the requested services from various applications. 

The rest of the paper is organized as follows. In Section~\ref{system}, we state our system model. In Section~\ref{Problem}, we formulate the intent profiling and translation problem. Our proposed framework solution is introduced in Section~\ref{framework}. Section~\ref{simulation} provides our simulation model and results. Finally, we conclude the paper in Section~\ref{conclusion}.

\section{System Model}~\label{system}

\begin{figure}
\centering
\includegraphics[width=3.5in,height=3in]{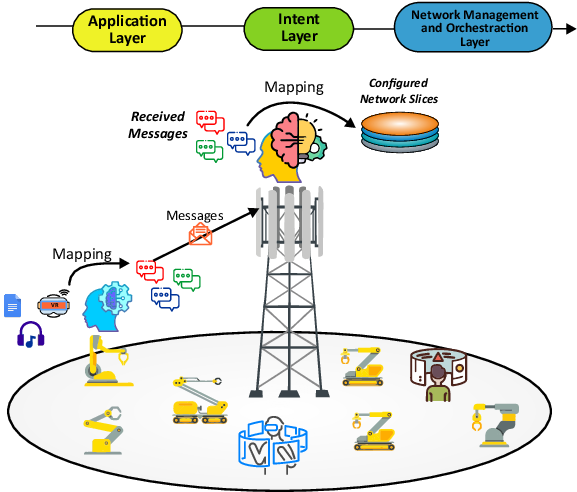}
\caption{System Model.}
\label{Fig2}
\end{figure}

We consider a network system consisting of~$N$ Industrial Internet of Things~(IIoT) mobile devices (MDs) indexed by~$\mathcal{N} =\{1,2,\dots,N\}$ and a single network equipped with~$M$ network slices indexed by~$\mathcal{M} =\{1,2,\dots,M\}$. The IIoT MDs are running different applications requesting different quality-of-experience (QoE) levels. The network is deployed with virtual network functions (VNF) and software-defined network (SDN) technologies to facilitate the implementation of network slices, where the network slices compose services with different capabilities using the network's communication, and computing resources, as shown in Fig.~\ref{Fig2}. The IIoT MDs communicate with the network over a time domain divided into time instances indexed by~$\mathcal{T} = \{1,2,\dots,T\}$. At each time instance~$t$, each IIoT MD $n$ runs an application and generates an intent instance denoted by~$I_{n,t}$  demanding a certain QoE. The requested QoE (i.e., intent) is then mapped to a QoS expressed in a maximum  communication and computation deadlines denoted by~$t^{\mathrm{up}}_{n,req}$ and $t^{\mathrm{comp}}_{n,req}$, respectively. 

 The network assigns a network slice~$m$ with capabilities~$(R_m,f_m)$ to each IIoT MD~$n$, where~$R_m$ is the uplink rate in bits per second, and~$f_m$ is the CPU computation resources in cycles per second. The supported data rate and computation resources from the allocated network slice must be sufficient to meet the communication and computation deadlines, otherwise, the requested QoE is not met. Based on the allocated network slice~$m$, the communication time (i.e., uplink time) of IIoT MD~$n$ can be calculated as
\begin{equation}
t^{\mathrm{up}}_n(m)= \frac{A_n}{R_m},      
\end{equation}
where~$A_n$ is the generated application instance task size of the IIoT MD~$n$. The computation time is computed as
\begin{equation}
t^{\mathrm{comp}}_n(m) = \frac{A_n \times C_n}{f_m},    
\end{equation}
where~$C_n$ is the number of required CPU cycles per bit of the IIoT MD-generated application task. 

The proposed framework operates over the following three layers: application, intent, and network management and orchestration. The application layer concerns the IIoT MDs and is responsible for profiling their intents. The intent layer communicates with IIoT MDs through ontology-based schemes and with the network through an interface. It is in charge of capturing the IIoT MDs' intent and translating it into network policies. The network management and orchestration layer is responsible for satisfying the requested QoE by applying the network policies through closed-loop control. The IIoT MDs communicate the requested QoE of their applications through intents in their domain language to the network, which the network needs to understand and map to the right capability. Due to the complexity of pre-defining the domain logic of each application on the network side, which is not practical nor scalable, this work investigates the decoupling of intent profiling and intent translation subproblems.

\section{Problem Formulation}\label{Problem}

We divide the intent profiling and translation problem into two subproblems. For the intent profiling subproblem, each IIoT MD communicates its intent (i.e., application instance requested QoE) to the network through a communication message chosen from a vocabulary set~$\mathcal{U}$.  The IIoT MDs aim to be associated with network slices that can satisfy their intents. Thus, each IIoT MD~$n$ optimizes a mapping function expressed as 
\begin{equation}\label{eq1}
f_n(I_{n,t}) : I_{n,t} \rightarrow  u_{n,t}  \; \; \forall \; n \in \mathcal{N}, \; t \in \mathcal{T}, \; u\in \mathcal{U}
\end{equation}
that maps each intent instance~$I_{n,t}$ to a communication message~$u_{n,t} \in \mathcal{U}$ to maximize the successful number of intents to communication messages association. For the intent translation subproblem, the network slices are deployed based on pre-defined network slice templates that can guarantee a certain QoS. We assume that the expressed intents are within the capabilities of the deployed network to guarantee QoS satisfaction. The network receives the communication messages from all IIoT MDs and optimizes a mapping function expressed as 
\begin{equation}\label{eq2}
g(\boldsymbol{u}_t) :  \boldsymbol{u}_t \rightarrow  \boldsymbol{c}_t
\end{equation}
that maps the communication messages to network slices to maximize the number of successful communication messages to network slices association, where~$\boldsymbol{u}_t \triangleq  [u_1,u_2,\dots,u_N]$ is a vector contains the received communication messages from all IIoT MDs at time slot~$t$. The vector $\boldsymbol{c}_t \triangleq  [c_{1},c_{2},\dots,c_{N}]$ contains the slices allocated to each IIoT MD at time slot~$t$, where~$c_{n} \in \mathcal{M}$ and the $N \times M$ binary association matrix at time slot~$t$ is therefore defined as $\boldsymbol{Y}^t \triangleq [y^t_{n,m}]$, indicates the association of network slices to IIoT MDs, where~$y_{n,m} \in \{0,1\}$ indicates that the network slice~$m$ is allocated to IIoT MD~$n$.

\textbf{\em The main objective of the system is to maximize the number of successful associations of intents to network slices that meet their requirements}. We define for each IIoT MD~$n$ an indicator~$x_{t,n}$ that takes value~$1$ in case of successful mapping between intents and network slices and~$0$ otherwise. We consider the mapping is successful if the allocated network slice~$m$ characteristics can satisfy the requested QoE, i.e,   
\begin{equation}
    x_{t,n} =
    \begin{cases}
        1 \;\; & \text{if } \; t^{\mathrm{comp}}_n \leq t^{\mathrm{comp}}_{n,req}\; \text{and} \; t^{\mathrm{up}}_n \leq t^{\mathrm{up}}_{n,req}, \\
        0 \;\; & \text{otherwise}.
    \end{cases}
\end{equation}
Subsequently, an optimization problem can be stated as follows
\begin{align*}
\max_{\boldsymbol{Y,u}} \; &  \sum_{t \in \mathcal{T}} \sum_{n\in \mathcal{N}} x_{t,n}  \\
\text{subject to} \\
&C1:\sum_{m\in \mathcal{M}} y^t_{n,m} = 1, \; \; \forall \;  n \in \mathcal{N}, \; t \in \mathcal{T}\\
&C2: x_{t,n} \in \{0,1\}.
\end{align*}

The first constraint indicates that each IIoT MD application instance can be associated with a maximum of one network slice. {\textbf{\em The main objective can be maximized through maximizing the objective functions Eq.~\eqref{eq1} and \eqref{eq2} that maximize the successful mapping between intents to communication messages and communication messages to network slices}}. Unfortunately, both mapping functions Eq.~\eqref{eq1} and \eqref{eq2} are unknown and hard to model explicitly mathematically. Thus we propose an AI framework in the next section to approximate and solve the problem.

\section{Proposed Framework}~\label{framework}

To solve the intent profiling and translation problem stated above, we propose an AI framework that utilizes cooperative multi-agent reinforcement learning (MARL), where the network and IIoT MDs are modeled as reinforcement learning agents. Moreover, we adopt emergent communication technology to learn a communication protocol that provides a
common ground between the intent expression and the
network capabilities. Emergent communication has been introduced as a way for AI agents to solve problems cooperatively or competitively through communication~\cite{foerster2016learning}. Communication emerges in the sense that the communication messages have no predefined meaning and through interaction, the agents assign meaning to them~\cite{lazaridou2020emergent}. Communication messages can take two forms, continuous or discrete. Although continuous communication can let the environment be represented as a single-agent network due to continuous back-propagation, it can propagate rich error information and cause poor performance. On the other hand, discrete communication messages, where agents send a symbol or sequence of symbols form a multi-agent environment and eliminate error propagation. Thus, in this work, we leverage discrete communication for intent profiling.

The MARL is described with a decentralized partially observable Markov decision process (Dec-POMDP) ~$\mathcal{P} = \langle \mathcal{S,O,A,T,R}, \gamma \rangle$, where~$\mathcal{S}$ is the state space,~$\mathcal{O}$ is the observation space,~$\mathcal{A}$~is the action space, $\mathcal{T}: \mathcal{S} \times \mathcal{A} \xrightarrow{} \Delta(S)$ is a non-deterministic transition function maps the state and action space to a probability distributions~$\Delta(\mathcal{S})$ over $\mathcal{S}$,  $R:\mathcal{S} \times \mathcal{A} \xrightarrow{} \mathbb{R}$ is a reward function, which maps the states and actions to a set of real numbers and~$\gamma$ is the discount factor. The agents communicate over an episode of maximum length~$T$ time instances. At time step~$t$, the agent receives an observation $o_t$ depending on its current state~$s_t$ and the previous action~$a_{t - 1}$ then takes an action.

The action space~$\mathcal{A}$ contains environment and communication actions. The environment action~$a_e$ represents the network slice allocation to each IIoT MD. The communication action~$a_c$ includes the uplink and downlink communication messages. The uplink messages~$U$ are chosen from the set~$\mathcal{U}$. The downlink messages ~$D \in \{0,1\}$ contain two messages indicating the success or failure of the network slice allocation in satisfying the requested QoE. Note that the uplink and downlink communication messages are not pre-defined and the meaning associated with each message emerges through communication. The network state space~$\mathcal{S}$ consists of the recent~$l$ uplink and downlink communication messages and the environment action
$\mathbf{S}^b = [U^n_t,\dots,U^n_{t-1-l},D^b_{t-1},\dots,D^b_{t-1-l},a^b_{t-1},\dots, a^b_{t-1-l}].$ Each IIoT MD state space is composed of the recent~$l$ uplink and downlink communication messages and the generated intent instance (i.e., application requested QoE) 
$\mathbf{S}^n = [U^n_{t-1},\dots,U^n_{t-1-l},D^b_{t-1},\dots,D^b_{t-1-l},I_{n,t},\dots, I_{n,t-1-l}]$. 
Since cooperative MARL is Dec-POMDP and the state of the environment is not available, we add an $l$ history to help the agents learn better. 
The reward at each time step is defined as
$$R_n(t) = \begin{cases}
+\rho & \text{ if the intent is satisfied}\\ 
-\rho & \text{otherwise} \\ 
\end{cases}$$
The reward is $+\rho$ if the IIoT MD intent is satisfied through the allocated network slice,~$-\rho$ otherwise. The team reward is the sum of the rewards of all IIoT MDs, which is defined as
$R(t) = \sum_{n \in \mathcal{N}} R_n(t).$

To solve the above formulated Dec-POMDP problem, we adopt the multi-agent proximal policy optimization (MAPPO) algorithm~\cite{yu2103surprising}, which is an extension from the PPO algorithm~\cite{schulman2017proximal} to solve cooperative tasks. Each agent architecture of MAPPO consists of two models, actor (i.e., policy) and critic (i.e., value). The idea behind MAPPO is that agents need to communicate and share information (i.e., observation, action, model parameters) during the sampling stage to solve the target task cooperatively. Then, during the learning stage, each agent applies the standard PPO training stage with a centralized value function (i.e., the input contains all agent's states) to compute the Generalized Advantage Estimation (GAE) and apply the PPO critic learning procedure. The actor decides the next action based on the current state while the critic evaluates the states. The actor update rule is done to optimize the surrogate-clipped objective function
$$L^{\mathrm{CLIP}}(\theta)  = \hat{\mathbb{E}}_t\big[\min(r_t(\theta)\hat{\boldsymbol{A}_t},\mathrm{clip}(r_t(\theta), 1 - \epsilon, 1 + \epsilon)\hat{\boldsymbol{A}_t}\big],$$
where~$\epsilon$ is a hyperparameter and~$r_t(\theta) = \frac{\pi_{\mathrm{\theta}}(a_t | o_t)}{\pi_{\mathrm{\theta}_{\mathrm{old}}}(a_t | o_t)}$ is the ratio between the new and old policies. The GAE~$\hat{\boldsymbol{A}_t}$ at each time step is calculated by 
$$\hat{\boldsymbol{A}_t} = \delta_t + (\gamma \lambda) \delta_{t +1} + \dots + \dots +  (\gamma \lambda)^{T -t +1} \delta_{T - 1},$$
where~$\delta_t = r_t + \gamma V^{\phi}(o_{t+1}) - V^{\phi}(o_t)$ counts the benefits of the new state over the old state. The centralized value function update rule is given by
$$\arg \min_{\phi} \frac{1}{|\mathcal{D}| T} \sum_{\tau \in \mathcal{D}} \sum^T_{t=0}  (V^{\phi}(o_t,s_t,\boldsymbol{a^-}) - \hat{R_t})^2$$
where~$\mathcal{D}$ is the collected trajectories from all agents,~$\tau$ is the trajectory, $\hat{R_t}$ is the rewards,~$\boldsymbol{a}$ is the actions of all agents except the current agent. The critic network is updated based on the collected experience from all agents (concatenate all agent's states as input to the critic network). 

\section{Simulation Model and Results}\label{simulation}

In this section, we evaluate the performance of our proposed framework. We consider a warehousing logistic area with a network and~$5$ IIoT MDs. The applications supported in the system are either ultra-reliable low latency communication (URLLC) or enhanced mobile broadband (eMBB). The network has ten network slices with different capabilities. The uplink messages set has a cardinality equal to the number of network slices supported on the system. The generated instances of the applications are generated based on the parameters listed in Table~\ref{table:1}. MAPPO is implemented with the hyperparameters listed in Table~\ref{table:2}, where the policy and value functions are represented by separate MLP fully connected linear neural networks and optimized by Adam optimizer~\cite{kingma2014adam}. The proposed framework is compared with the following baselines:

\begin{itemize}
\item \textbf{Perfect Knowledge:} The network checks all possible network slices available that can guarantee the requested QoE to IIoT MD (i.e., intent). Then, it chooses one of them at random.
\item \textbf{Random Assignment:} The network allocates network slices to the IIoT MDs (i.e., intents) in a random way.
\item \textbf{Self-Learning Slice Selection:} The IIoT MDs learn to access the network slices without any prior assignment or communication with the network through interaction via RL. 
\end{itemize}

\begin{table}
\caption{Simulation Parameters}
\centering
\begin{tabular}{|c|c|}
\hline
\textbf{Parameters} & \textbf{Values}\\
\hline
No. of IIoT MDs  & $5$ \\
\hline
No. of network slices  & $10$ \\
\hline
Tasks Size & $100-500$ bits\\
\hline
Tasks Computation Requirement  & $1 \times 10^2 - 5 \times 10^4$\\
\hline
Tasks Storage Requirement  & $200 - 600$ bits\\
\hline
IIoT MDs Reliability Requirement  & $1 \times 10^{-2} - 5 \times 10^{-5}$\\
\hline
Tasks Offloading Tolerance & $1 \times 10^{-2} - 5 \times 10^{-2}$~second\\
\hline
Tasks Computation Tolerance & $1 \times 10^{-2} - 5 \times 10^{-2}$~second\\
\hline
Probability of Task Arrival & $1$\\
\hline
Duration of episode & 15\\
\hline
\end{tabular}
\label{table:1}
\end{table}

\begin{table}
\caption{MAPPO Hyperparameters}
\centering
\begin{tabular}{|c|c|c|c|}
\hline
$\textbf{Hyperparameter}$ & $\textbf{Values}$ & $\textbf{Hyperparameter}$ & $\textbf{Values}$\\
\hline
Number of episodes & 6000 & Learning rate & $10^{-3}$ \\
\hline
Minibatch size & $64$ & Discount factor~$(\gamma)$& $0.99$\\
\hline
GAE parameter~$(\lambda)$ & $0.95$ & Clipping parameter~$(\epsilon)$ & $0.2$\\
\hline
VF coeff.~$(c1)$ & $0.2$ & Entropy coeff.~$(c2)$ & $0.2$ \\
\hline
Optimizer & Adam & Optimizer epsilon & $10^{-5}$\\
\hline
\end{tabular}
\label{table:2}
\end{table}

\begin{figure}
    \centering
    \includegraphics[width = 3.5 in, height = 2.3 in]{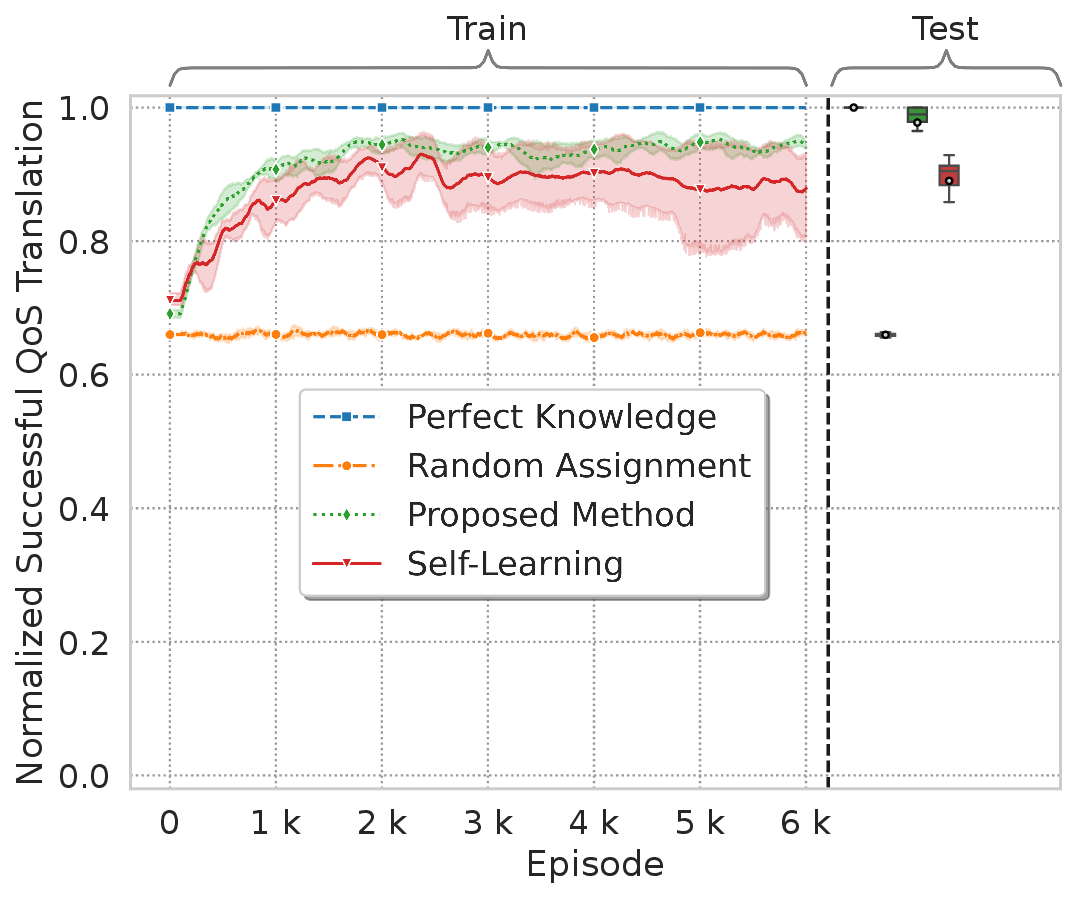}
    \caption{Normalized successful QoS translations versus the number of episodes.}
   \label{Fig3}
\end{figure}

\begin{figure}
    \centering
    \includegraphics[width = 3.5 in, height = 2.3 in]{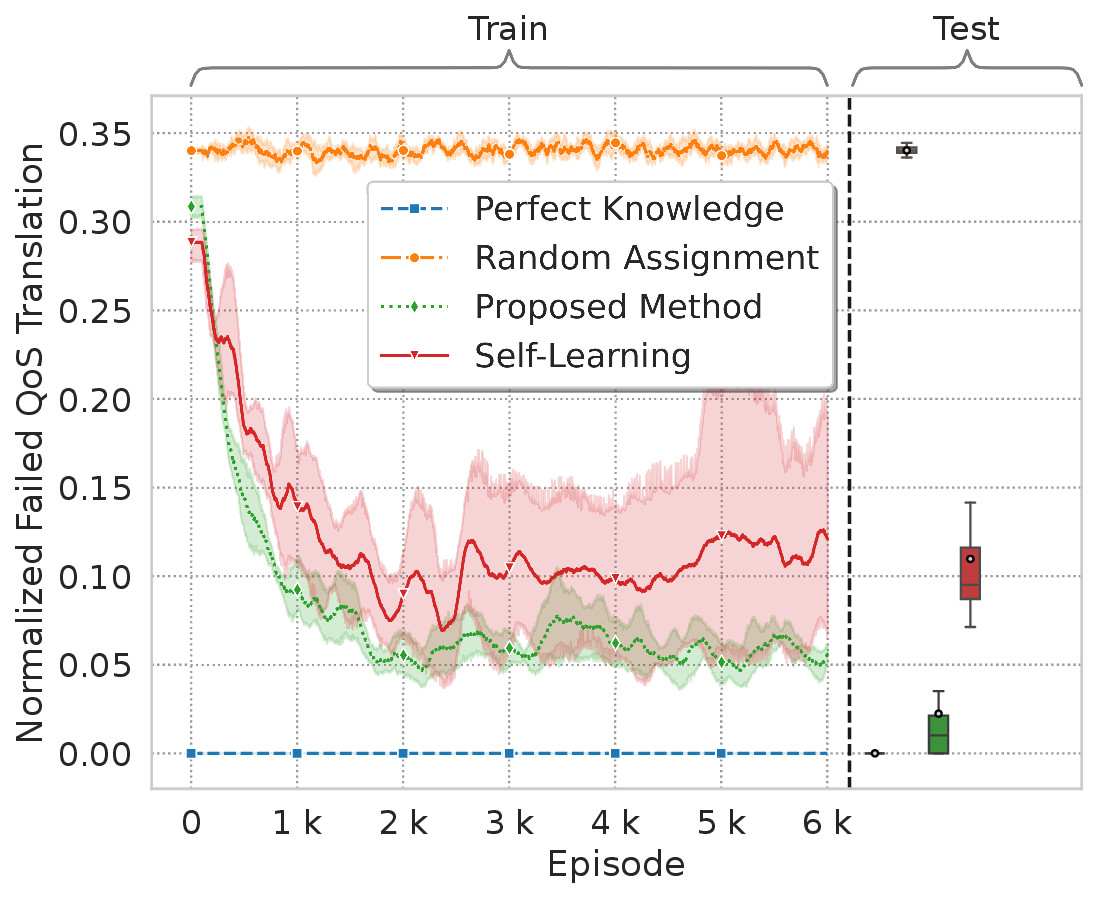}
    \caption{Normalized failed QoS translations versus number of episodes.}
   \label{Fig4}
\end{figure}

Fig.~\ref{Fig3} demonstrates the normalized successful translated intents versus the number of episodes, which reflects the success of the network in allocating the network slices that can satisfy the IIoT MD intent correctly based on the communicated messages. As we can observe, the proposed scheme outperforms the random assignment and self-learning selection schemes and reaches a very close performance to the perfect knowledge approach. Moreover, it gives an outstanding performance during the testing phase. Fig.~\ref{Fig4} shows the normalized failed intent translations versus the number of episodes. The proposed intent profiling and translation scheme reduces the number of failed translations compared to the random assignment and self-learning selection strategies. Furthermore, the proposed scheme approaches zero failure in the testing phase.

\begin{figure}
    \centering
    \includegraphics[width = 3.5 in, height = 2.3 in]{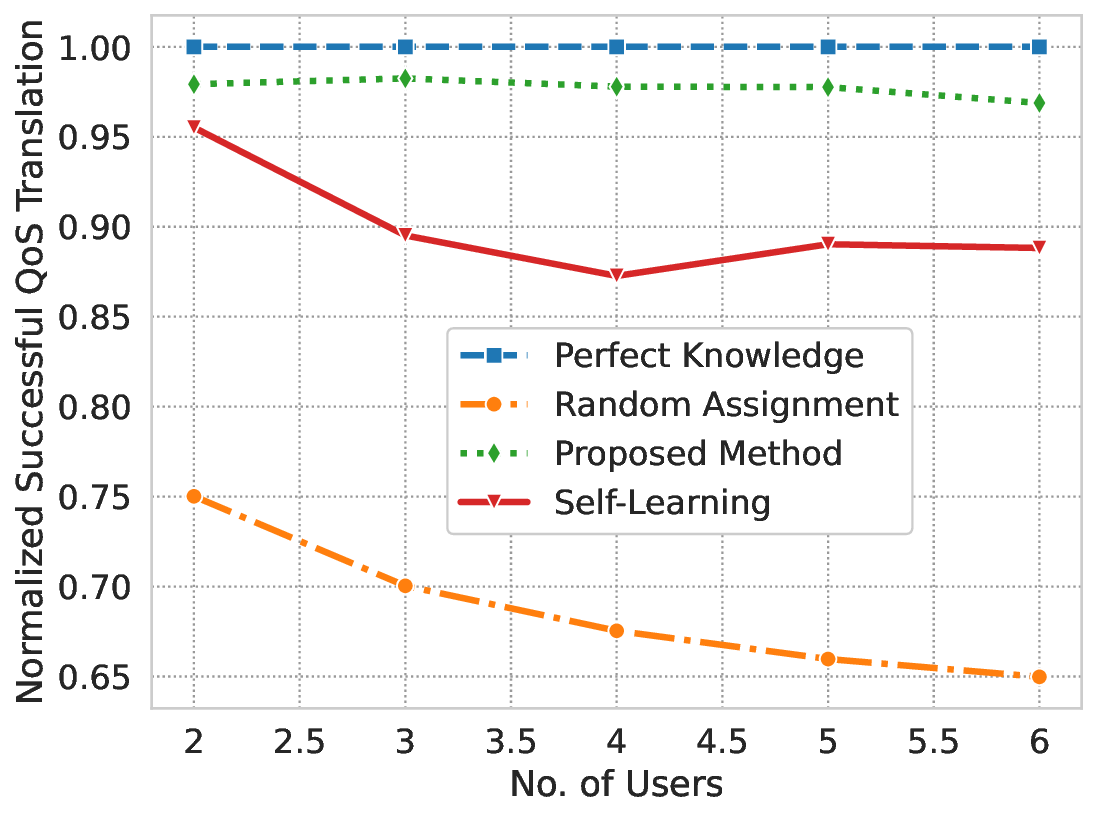}
    \caption{Normalized successful QoS translations versus the number of users.}
   \label{Fig5}
\end{figure}

\begin{figure}
    \centering
    \includegraphics[width = 3.5 in, height = 2.3 in]{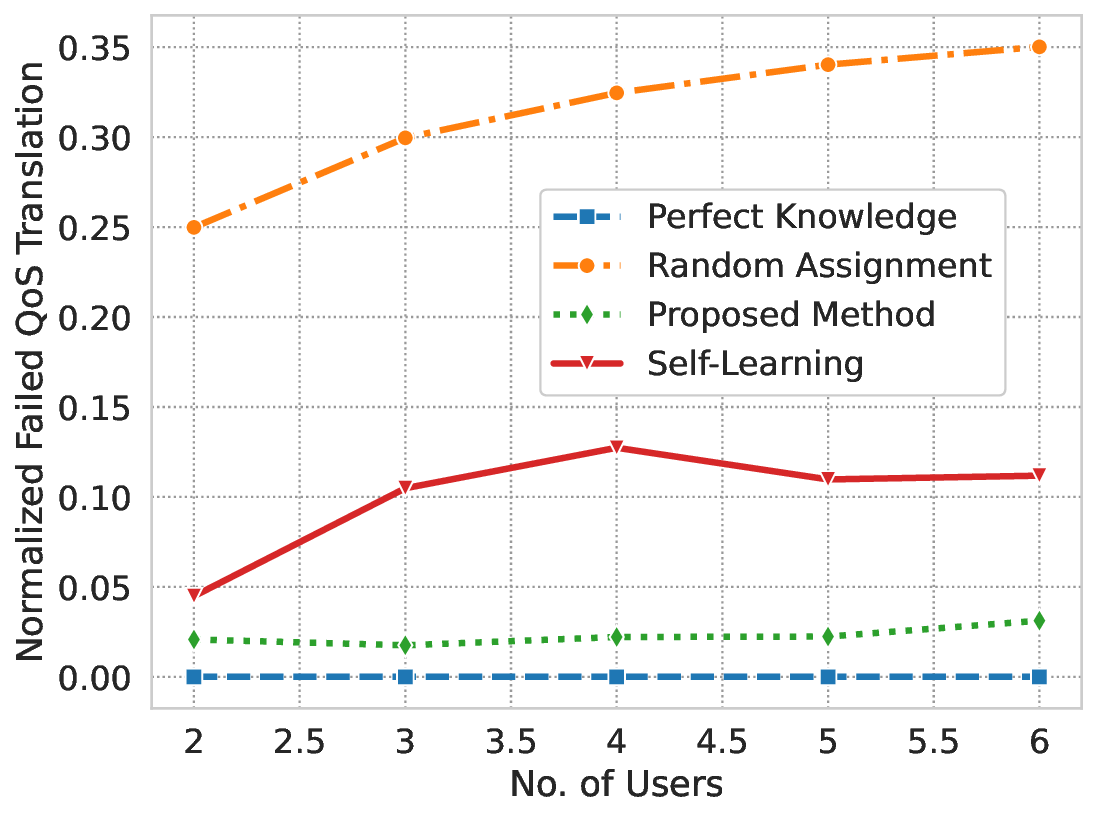}
    \caption{Normalized failed QoS translations versus the number of users.}
   \label{Fig6}
\end{figure}

Fig.~\ref{Fig5} shows the normalized successful translated intents versus the number of users. Obviously, as the number of users increases, the number of successful QoS translations increases. As we can see, the proposed scheme gives a very close performance to the perfect knowledge approach. However, as the number of users increases, the gap between them increases as well which is an expected performance due to the large number of messages received at the network to interpret. Fig.~\ref{Fig6} demonstrates the normalized failed intent translations versus the number of users. As we can notice, as the number of users increases, the number of failed QoS translations increases for the same aforementioned reason. 

\section{Conclusion}\label{conclusion}

We investigated the problem of intent profiling and translation to provide a simple, efficient, and automated way to manage and operate intent-based networks. The proposed scheme leverages machine learning and emergent communication, where the IIoT MDs (i.e., applications) learn a policy to map their intents to communication messages to the network. Afterward, the network learns a policy to translate these messages to network resources (i.e., network slice allocation). The proposed scheme outperformed the random assignment and self-learning selection strategies during training and testing. It also gives a very close performance to the perfect knowledge benchmark scheme. 

\section{ACKNOWLEDGMENT}

The work is funded by the project SCENE (G.A no. 00164501.0). The work is also funded by the European Union through the projects 6G-INTENSE (G.A no. 101139266), CENTRIC (G.A no. 101096379), and VERGE (G.A no. 101096034). Views and opinions expressed are however those of the author(s) only and do not necessarily reflect those of the European Union. Neither the European Union nor the granting authority can be held responsible for them. This research was supported by the Research Council of Finland (former Academy of Finland) 6G Flagship Programme (Grant Number: 346208). 

\bibliographystyle{ieeetr}
\bibliography{References}

\end{document}